\documentclass[aps,showpacs]{revtex4}
\usepackage{epsfig,graphicx}
\usepackage{amsmath,amsfonts}

\begin{document}

\title{Self-force of a point charge in the space-time of a massive wormhole}

\author{Nail R. Khusnutdinov\footnote{e-mail: 7nail7@gmail.com}${}^{a}$,
Arkady A. Popov\footnote{e-mail: apopov@ksu.ru}${}^{b}$and \ Lilia N. Lipatova$^{a}$
}

\affiliation{${}^a$Department of Physics, Kazan State University, Kremlevskaya 18,
Kazan 420008, Russia \\ ${}^b$Department of Physics and Mathematics Education,
Tatar State University of Humanity and Pedagogic, Tatarstan 2, Kazan 420021, Russia}


\begin{abstract}
We consider the self-potential and the self-force for an
electrically charged particle at rest in the massive wormhole
space-time. We develop general approach for the renormalization of
electromagnetic field of such particle in the static space-times
and apply it to the space-time of wormhole the metric of
which is a solution to the Einstein-scalar field equations in the
case of a phantom massless scalar field. The self-force is found
in manifest form; it is an attractive force. We discuss the
peculiarities due to the parameter of the wormhole mass.
\end{abstract}

\maketitle

\section{Introduction}

The wormhole is an example of the space-time with nontrivial topology and it
connects different  universes or different parts of the same universe. Careful
discussion of wormhole geometry and physics as well as a good introduction in
the subject may be found in the Visser book \cite{Vis95} and in the review by
Lobo \cite{Lob08}. The most recent growth of interest in wormholes was connected with the "time machine", introduced by Morris, Thorne, and Yurtsever in Refs. \cite{MorTho88,MorThoYur88}.

In curved spacetimes there is an interesting interaction charged particle with its
own electromagnetic field due to scattering on the curvature \cite{DeWBre60}.
The origin of this self-force is associated with nonlocal structure of the
field, the source of which is the particle. For particles in the spacetime of a
cosmic string \cite{Lin86}, the self-force may be the unique local gravitational
interaction on a particle. Therefore, in this case there is no local
gravitational interaction between the particle and cosmic string, but
nevertheless there exists a repulsive self-interaction force which is of
nonlocal origin. A discussion of the self-force in detail may be found in reviews
\cite{Poi04,Khu05}.

In calculating the self-force acting on a particle one typically encounters
a divergent expression. The finite expression for the self-force should be obtained by using a certain renormalization prescription. In the present paper we
develop a simple renormalization procedure for vector potential which may be applied for charged particle at rest in static space-times. The main idea
of this renormalization method originates from the well-developed procedure
of renormalization of the Green's functions in quantum field theory on
the background of curved space-times. In framework of this procedure one subtracts some terms of expansion of corresponding Green function of massive field from the divergent expression obtained. The quantities of terms to be subtracted is defined by simple rule -- they no longer vanish as the field's mass  goes to infinity. This approach has been developed in Refs.  \cite{1Roth2004,2Roth2004} for the massless scalar field.

The self-force for a charged particle at rest in the static wormhole background was analysed in detail in Refs. \cite{KhuBak07,Lin08,Kra08,BezKhu09} for arbitrary and specific profiles of the throat. The space-time was a spherical symmetric solution the Einstein equations with phantom scalar field as a source. Sometimes this background is called anti-Fisher due to analogous solution obtained by Fisher \cite{Fis48} for normal scalar field. The space-time of wormhole has had ultrastatic form, $g_{tt} = 1$, and therefore the Newtonian potential equals to zero. For this reason we call this background as massless wormhole. It was shown that for electric charge the self-force is always attractive \cite{KhuBak07,Lin08,Kra08} whereas it may be attractive or repulsive for scalar charge  \cite{BezKhu09} depending on the non-minimal connection constant $\xi$. In the present paper we analyse the self-force for electric particle at rest in space-time of another kind of wormhole considered in the Refs. \cite{Bron73,Ell73} with phantom scalar field as a source. We call this space-time as massive wormhole because the Newtonian potential is not zero (see Sec. \ref{Sec:Space-Time}).

The organization of this paper is as follows. In Sec. \ref{Sec:Space-Time}, we consider the space-time properties of the background under consideration. In the Sec. \ref{Sec:Maxwell} we obtain in manifest form solution of the Maxwell equations for particle at rest in the background of the massive wormhole \cite{Bron73,Ell73}. The Sec. \ref{Sec:Renorm} is devoted to renormalization procedure for electromagnetic field of such particle in  the case of static space-times. The procedure suggested is applied
for the well-known case the Schwarzschild black hole. In the Sec \ref{Sec:Self} we calculate in manifest form the self-energy and self-force for the electrically charged particle in the space-time of the massive wormhole \cite{Bron73,Ell73}. We discuss the results in Sec. \ref{Sec:Conclusion}. Our conventions are those of Misner, Thorne, and Wheeler \cite{MTW73}. Throughout this paper, we use units $c=  G=  1$.

\maketitle

\section{The space-time}\label{Sec:Space-Time}
The line element of massive wormhole has the following form \cite{Bron73,Ell73}
\begin{equation}\label{metric}
ds^2 = -e^{-\alpha(\rho)} dt^2 + e^{\alpha(\rho)}d\rho^2 + r^2(\rho) d\Omega^2,
\end{equation}
where
\begin{eqnarray}
r^2(\rho) &=& p^2 e^{\alpha(\rho)},\\
\alpha(\rho) &=& \frac{2m}{\sqrt{n^2-m^2}} \left\{ \frac{\pi}{2}
- \arctan \left(\frac{\rho}{\sqrt{n^2-m^2}}\right) \right\},
\end{eqnarray}
and $p^2 = \rho^2 + n^2 -m^2$. The radial coordinate $\rho$ may be positive as
well as negative, too.  The square of the sphere of radial coordinate $\rho$,
$S=4\pi r^2(\rho)$, is minimized for $\rho = m$.

There are two parameters, $n$ and $m$, in the metric (\ref{metric}) with relation $m^2 < n^2$. We consider the case $m>0$ without the loss of generality. For $m=0$ we obtain the wormhole which was called as massless "drainhole" \cite{Ell73} with throat
radius $n$. To reveal the role of the parameter $m$ let us consider the
gravitational acceleration, $a$, acting for a particle. In spherical symmetric case it has the well-known form \cite{LanLifV2,DorKarNovNov08}
\begin{equation}
a=\sqrt{g^{\rho\rho}} \, a_{\rho} = -\frac{1}{\sqrt{g_{\rho\rho}}}\frac{d\ln{\sqrt{-g_{tt}}}}{d\rho}.
\end{equation}
For the metric under consideration (\ref{metric}) we obtain
\begin{equation}
a = -\frac{m}{p^2}e^{\alpha/2}.
\end{equation}
In the limit $\rho\to +\infty$ we have the Newtonian gravitational attraction, $a\approx-{m}/{\rho^2}$,  to the wormhole throat with effective mass $m$. For $\rho\to -\infty$ we obtain Newtonian gravitational repulsion,
$a\approx-{me^{\alpha_{max}/2}}/{\rho^2}$, with negative effective mass
$-me^{\alpha_{max}/2}$, where $\alpha_{max} = \alpha_{\rho\to -\infty}$.
Therefore the parameter $m$ characterizes the effective mass of the wormhole.

The parameter $n$ of the wormhole may be understand in terms of a charge some massless scalar field going through the wormhole (for electric charge, see Ref. \cite{ShaNovKar08}). Indeed, the equation for scalar massless field in the background with line element (\ref{metric}) admits radial non-singular solution for scalar field.  The derivative of this field with
respect of the radial coordinate reads $\phi'(\rho) = -q/p^2$ with some
parameter $q$ which corresponds to the scalar charge from the point of view
of a distant observer. The stress-energy tensor of this field has
the following form $8\pi T^\nu_\mu  = q^2 e^{-\alpha}/p^4
\textrm{diag}(-1,1,-1,-1)$. On the other hand, the metric (\ref{metric})
is a non-singular solution of the Einstein equations with scalar
massless field with opposite sign of the stress-energy tensor which reads
$8\pi T^\nu_\mu = -n^2 e^{-\alpha}/p^4 \textrm{diag}(-1,1,-1,-1)$.
Therefore, the parameter $n$ plays the role of the scalar charge
of the phantom scalar field which pass through the wormhole space-time.

\section{The Maxwell equations}\label{Sec:Maxwell}
The Maxwell equations under the covariant Lorentz gauge have the following form ($e$ is a electric charge of the particle)
\begin{equation}
g^{\alpha\beta} A_{\mu;\alpha\beta} - A_\nu R^\nu_\mu = -4\pi J_\mu = -4\pi e \int u_\mu (\tau) \delta^{(4)}(x-x(\tau))\frac{d\tau}{\sqrt{-g}}.
\end{equation}
In manifest form they read
\begin{eqnarray*}
-4\pi e^{\alpha} J_t &=& A_{t,\rho\rho} + \frac{2(\rho - m)}{p^2}A_{t,\rho} +
\frac{1}{p^2} \hat{L}^2 A_t - e^{2\alpha} A_{t,tt} + \frac{2m}{p^2} A_{\rho,t}, \allowdisplaybreaks\\
-4\pi e^{\alpha} J_\rho &=& A_{\rho,\rho\rho} + \frac{2(\rho + m)}{p^2} A_{\rho,\rho} + \frac{1}{p^2} \hat{L}^2 A_\rho - \frac{2(m^2 -n^2 -2 m \rho + \rho^2)}{p^4} A_{\rho} - e^{2\alpha} A_{\rho,tt}\nonumber\allowdisplaybreaks\\
&-& \frac{2(\rho - m)}{p^4} A_{\theta,\theta} - \frac{2(\rho - m)\cot \theta}{p^4} A_{\theta} - \frac{2(\rho - m)}{p^4\sin^2\theta} A_{\varphi,\varphi} + \frac{2me^{2\alpha}}{p^2} A_{t,t}, \allowdisplaybreaks\\
-4\pi e^{\alpha} J_\theta &=& A_{\theta,\rho\rho} + \frac{2m}{p^2}A_{\theta,\rho} + \frac{1}{p^2} \hat{L}^2 A_\theta - \frac{1}{p^2\sin^2\theta } A_\theta - e^{2\alpha} A_{\theta,tt} + \frac{2(\rho - m)}{p^2} A_{\rho,\theta}  - \frac{2\cot\theta }{p^2\sin^2\theta } A_{\varphi,\varphi}, \allowdisplaybreaks\\
-4\pi e^{\alpha} J_\varphi &=& A_{\varphi,\rho\rho} + \frac{2m}{p^2}A_{\varphi,\rho} + \frac{1}{p^2} \hat{L}^2 A_\varphi - \frac{2\cot\theta}{p^2} A_{\varphi,\theta} - e^{2\alpha} A_{\varphi,tt} + \frac{2(\rho - m)}{p^2} A_{\rho,\varphi}  + \frac{2\cot\theta }{p^2} A_{\theta,\varphi},
\end{eqnarray*}
and the Lorentz condition is
\begin{equation}
A_{\rho,\rho} + \frac{2\rho}{p^2} A_\rho + \frac{1}{p^2} A_{\theta,\theta} +  \frac{\cot\theta}{p^2} A_\theta + \frac{1}{p^2 \sin^2\theta} A_{\varphi,\varphi} - e^{2\alpha} A_{t,t} = 0,
\end{equation}
where $p^2 = \rho^2 + n^2 - m^2$, and $\hat{L}^2 = \partial^2_{\theta} + \cot \theta \partial_{\theta} + \csc^2 \theta \partial^2_{\varphi}$.

For particle at rest with position $x'$ the vector potential $A_\mu$ has no
dependence  on the time and we use the following anzatz: $A_\mu = (0,0,0,A_t)$.
The equation for Lorentz gauge is identically satisfied and the system of
equations reduces for single equation for $A_t$ \footnote{We use covariant
component of the potential in contrast with the paper \cite{KhuBak07} and for
this reason there is no sign minus here in right hand side.} :
\begin{equation}\label{at}
A_{t,\rho\rho} + \frac{2(\rho - m)}{p^2}A_{t,\rho} +
\frac{1}{p^2} \hat{L}^2 A_t  = \frac{4\pi e}{e^{\alpha}p^2\sin\theta} \delta^{(3)} (x - x').
\end{equation}

Due to spherical symmetry of the problem under consideration we represent the potential in the following form
\begin{equation}
A_t = 4\pi e \sum_{l,m} Y_{lm}(\Omega)Y^*_{lm}(\Omega') g_l(\rho,\rho'),
\end{equation}
where $Y_{lm}(\Omega)$ is the spherical functions of argument $\Omega = (\theta, \varphi)$. The radial part, $g_{l}$, satisfies the equation
\begin{equation}
g_{l,\, \rho \rho} + \frac{2(\rho - m)}{p^2}g_{l, \, \rho} - \frac{l(l+1)}{p^2} g_l =
\frac{1}{e^{\alpha}p^2} \delta(\rho - \rho').
\end{equation}

To solve this equation we use approach developed in Ref. \cite{KhuBak07}.
Namely, the solution  of the above equation is represented in the following form
\begin{equation}
g_{l}(\rho,\rho') = \theta (\rho - \rho') \Psi_1 (\rho') \Psi_2 (\rho) + \theta (\rho' - \rho) \Psi_1 (\rho) \Psi_2 (\rho'),
\end{equation}
where $\Psi_1$ and $\Psi_2$ are the increasing and decreasing for $\rho\to
\infty$ solutions  of the corresponding homogeneous equation with Wronskian
$W(\Psi_1,\Psi_2) = -e^{-\alpha}/p^2$.  These solutions in its turn are
expressed in terms of the full set of solutions of corresponding homogeneous
equation for both domain of the space namely, for $\rho >0$ and $\rho <0$. The
full set of solutions read
\begin{subequations}
\begin{eqnarray}
\phi_1^+ &=& c_1^+ e^{b \arctan x} P_l^{ib}(ix),\ \rho >0, \allowdisplaybreaks \\
\phi_2^+ &=& c_1^+ e^{b \arctan x} Q_l^{ib}(ix),\ \rho >0, \allowdisplaybreaks \\
\phi_1^- &=& c_1^+ e^{b \arctan x} P_l^{ib}(-ix),\ \rho <0, \allowdisplaybreaks \\
\phi_2^- &=& c_1^+ e^{b \arctan x} Q_l^{ib}(-ix),\ \rho <0, \allowdisplaybreaks
\end{eqnarray}
\end{subequations}
where
\begin{equation}
x = \frac{\rho}{\sqrt{n^2 - m^2}},\ b = \frac{m}{\sqrt{n^2 - m^2}},
\end{equation}
and $P_l^{ib},\ Q_l^{ib}$ are the modified Legendre functions of the first  and the second kind \cite{BatErdV1}. Then we have to consider the 6 different relations $\rho$ and $\rho'$ \cite{KhuBak07} depending on the sign and value of radial coordinate. To calculate self-force we need only for $\rho > \rho'> 0$ and $0 > \rho' > \rho$.  Taking into account the Wronskian (see Ref. \cite{BatErdV1}),
\begin{equation}
W(P_l^{\mu}(z),Q_l^{\mu}(z)) = \frac{e^{i\pi\mu}}{1-z^2} \frac{\Gamma (1+\nu+\mu)}{\Gamma (1+\nu - \mu)},
\end{equation}
we obtain
\begin{eqnarray*}
A_t^{+} &=& -\frac{e e^{b \arctan x + b \arctan x'}}{\sqrt{n^2 - m^2}}   \sum_{l=0}^\infty (2l+1) \left\{i e^{-2\pi b} P_l^{ib} (ix') Q_l^{-ib} (ix) + \frac{1}{\pi} Q_l^{ib} (ix') Q_l^{-ib} (ix)\right\}, \allowdisplaybreaks \\
A_t^{-} &=& -\frac{e e^{b \arctan x + b \arctan x'}}{\sqrt{n^2 - m^2}}   \sum_{l=0}^\infty (2l+1) \left\{i e^{-2\pi b} P_l^{ib} (-ix') Q_l^{-ib} (-ix) + \frac{1}{\pi} Q_l^{ib} (-ix') Q_l^{-ib} (-ix)\right\},
\end{eqnarray*}
where $A_t^{+}$ is the vector potential for $\rho > \rho'> 0$ and $A_t^{-}$ is for $0 > \rho' > \rho$.

To find expressions in close form for the series above  we take into account the Heine formula
\begin{equation}
\sum_{l=0}^\infty (2l+1) P_l (x') Q_l(x) = \frac{1}{x-x'},
\end{equation}
and integral representations for the Legendre functions \cite{BatErdV1}
\begin{eqnarray}
P_l^\mu (x') &=& \frac{(x'^2-1)^{\mu/2}}{\Gamma (-\mu)} \int_1^{x'} P_l(\tau) (x'-\tau)^{-\mu-1},\ \Re\mu >0,x'\not \in [-1,1],\allowdisplaybreaks\\
Q_l^{-\mu} (x) &=& e^{i\pi\mu} \frac{(x^2-1)^{-\mu/2}}{\Gamma (\mu)} \int_x^\infty Q_l(t) (t-x)^{\mu-1},\ l+1>\Re\mu>0, |x|>1.
\end{eqnarray}
Then after some machinery we obtain
\begin{equation}
\sum_{l=0}^\infty (2l+1) P_l^{ib} (ix') Q_l^{-ib}(ix) = -\frac{i}{x-x'} e^{\pi b + b \arctan x - b \arctan x'}
\end{equation}
for $x > x' > 0$ or $x < x' < 0$.
Taking into account the formula
\begin{equation}
Q_l^{ib}(ix') = \frac{\pi}{2} \frac{ie^{-\pi b}}{\sinh \pi b} \left\{(-1)^l P_l^{ib} (-ix') - P_l^{ib} (ix')\right\}
\end{equation}
we obtain another series in close form
\begin{equation}
\sum_{l=0}^\infty (2l+1) Q_l^{ib} (ix') Q_l^{-ib}(ix) = \frac{e^{-b \arctan x + b \arctan x'} - e^{b \arctan x - b \arctan x'}}{x-x'} \frac{\pi}{2\sinh \pi b}, \ \forall x,x'.
\end{equation}

Using these formulas we obtain
\begin{eqnarray}
A_t^{+} &=& -\frac{e}{\rho -\rho'} e^{-\alpha(\rho)} + \frac{e e^{\pi b}}{2 \sinh \pi b} \frac{e^{-\alpha(\rho)} - e^{-\alpha(\rho')}}{\rho - \rho'},\ \rho>\rho'>0, \allowdisplaybreaks\\
A_t^{-} &=& -\frac{e}{\rho' -\rho} e^{-\alpha(\rho)} + \frac{e e^{\pi b}}{2 \sinh \pi b} \frac{e^{-\alpha(\rho)} - e^{-\alpha(\rho')}}{\rho - \rho'},\ \rho <\rho'<0.
\end{eqnarray}
As noted above, these expressions are divergent at $\rho' \rightarrow \rho$,
and should be renormalized.

\section{Renormalization}\label{Sec:Renorm}

We adopt here a general approach to renormalization in curved space-time \cite{BirDav82}, which means the subtraction of the first terms of the DeWitt-Schwinger asymptotic expansion of the Green's function. The renormalization procedure for the space-time with line element (\ref{metric}) is not so simple as it was for massless wormhole
\cite{KhuBak07}. The point is that the space-time under consideration has no 
ultrastatic form, that is $g_{tt} \not = 1$. For this reason the
equation for $A_t$ in static case does not coincide with that for
scalar $3D$ Green function and we can not use the standard
formulas for DeWitt-Schwinger expansion of the $3D$ Green
function. Let us simplify the problem of renormalization by reduction
them to $3D$ case for the potential $A_t$ of charged particle at rest
in a static spacetime
         \begin{equation}
         ds^2= g_{tt}(x^i)d t^2+g_{j k}(x^i)d x^j d x^k,
         \end{equation}
where $i, j, k = 1, 2, 3$. This means that
the equation for Lorentz gauge is identically satisfied and
one can write out the equation for the single nonzero component of $A_{\mu}$
in the case of massive field in the following way
\begin{equation} \label{fem}
g^{\mu \nu } A_{t; \mu \nu} - A_t R^t_t-m_{\mbox{\tiny ef}}^2 A_t
= -4\pi e \int u_t (\tau)
\delta^{(4)}(x, x'(\tau))\frac{d\tau}{\sqrt{-g}}
=4 \pi e\frac{\sqrt{-g_{tt}}}{ \sqrt{g^{(3)}}}\, \delta^{(3)}(x^i, {x^i}')
\end{equation}
or
\begin{eqnarray}
g^{i j } \frac{\partial^2 A_t}{\partial x^i \partial x^j}
-\left( g^{i j} \Gamma^k_{ij}
+\frac{g^{i k}}{2 g_{tt}} \frac{\partial g_{tt}}{ \partial x^i}
\right) \frac{\partial A_t}{\partial x^k}
&&
+g^{i j} \left(
-\frac{\partial^2 g_{tt}}{2 g_{tt} \partial x^i \partial x^j}
+\frac{1}{4 g_{tt}^2}\frac{\partial g_{tt}}{ \partial x^i }
\frac{\partial g_{tt}}{\partial x^j }
+\Gamma^k_{ij} \frac{\partial g_{tt}}{2 g_{tt} \partial x^k }
 \right) A_t
 \nonumber \\ &&
 - A_t R^t_t-m_{\mbox{\tiny ef}}^2 A_t
=4 \pi e\frac{\sqrt{-g_{tt}}}{ \sqrt{g^{(3)}}}\, \delta^{(3)}(x^i, {x^i}'),
\end{eqnarray}
where $m_{\mbox{\tiny ef}}$ is mass of the field, $g^{(3)}=\det
g_{i j}$  and we have taken into account that $d\tau/dt=\sqrt{-g_{tt}}$.
From this equation we obtain the following equation for tetrad
component of the vector field $A_{(t)} =  A_t/\sqrt{-g_{tt}}$:
        \begin{eqnarray} \label{beq0}
        g^{i j } \left( \frac{\partial^2 A_{(t)}}{\partial x^i \partial x^j}
        - \Gamma^k_{ij}  \frac{\partial A_{(t)}}{\partial x^k}\right)
        -m_{\mbox{\tiny ef}}^2 A_{(t)}
        +\frac{g^{j k}}{2 g_{tt}}\frac{\partial g_{tt} }{\partial x^j}
        \frac{\partial A_{(t)}}{\partial x^k}
        -\left(\frac{g^{i j}}{4 g_{tt}^2}\frac{\partial g_{tt}}{ \partial x^i }
        \frac{\partial g_{tt}}{\partial x^j }
        +R^t_t \right) A_{(t)}
        =4 \pi e \frac{\delta^{(3)}(x^{i}, {x^i}')}{\sqrt{g^{(3)}}}.
        \end{eqnarray}
In the case of $m_{\mbox{\tiny ef}} \gg 1 / L$ we can obtain the solution
of this equation near the point $x'$ in terms of the distance along
the geodesic between the separated points and purely geometrical
quantities constructed out of the Riemann tensor \cite{DW,BP}.
Let us consider the equation for the
three-dimensional Euclidean Green's function $G_{\mbox{\tiny
E}}(x^{i}, {x^{i}}')$
        \begin{eqnarray} \label{beq}
        \frac{1}{\sqrt{g^{(3)}}} \frac{\partial}{\partial x^j}
        \left( \sqrt{g^{(3)}} g^{j k} \frac{\partial
        G_{\mbox{\tiny E}}(x^{i},{x^i}')}{\partial x^k}   \right)
        -m_{\mbox{\tiny ef}}^2 G_{\mbox{\tiny E}}(x^{i},{x^i}')
        +\frac{g^{j k}}{2 g_{tt}}\frac{\partial g_{tt} }{\partial x^j}
        \frac{\partial G_{\mbox{\tiny E}}
        (x^{i},{x^i}')}{\partial x^k} \nonumber \\
        -\left(
        \frac{g^{i j}}{4 g_{tt}^2}\frac{\partial g_{tt}}{ \partial x^i }
        \frac{\partial g_{tt}}{\partial x^j }
        +R^t_t \right) G_{\mbox{\tiny E}}(x^{i},{x^i}')
        =-\frac{\delta^{(3)}(x^{i}, {x^i}')}{\sqrt{g^{(3)}}}
        \end{eqnarray}
and introduce the Riemann normal coordinates $y^i$ in 3D space with
origin at the point ${x^i}'$ \cite{Pet}. In these coordinates one
has
        \begin{equation}
        g_{i j}(y^i)=\delta_{i j}
        -\frac13 { \widetilde R_{i k j l}}|_{y=0}\, y^k y^l+ O\left(\frac{y^3}{L^3}\right),
        \end{equation}
        \begin{equation}
        g^{i j}(y^i)=\delta^{i j}
        +\frac13 {\widetilde R^{i \ j}_{\ k \ l}}|_{y=0}
        \, y^k y^l+  O\left(\frac{y^3}{L^3}\right),
        \end{equation}
        \begin{equation} \label{exg}
        g^{(3)}(y^i)=1-\frac13 {\widetilde R_{i j}}|_{y=0} y^i y^j
        +  O\left(\frac{y^3}{L^3}\right),
        \end{equation}
where $\delta_{i j}$ denotes the metric of
a flat three-dimensional  Euclidean spacetime and $L$ is the
characteristic curvature scale of background geometry. $\widetilde
R_{i k j l}$ and $\widetilde R_{i j}$ stand for the components of
Riemann and Ricci tensors of the three-dimensional spacetime with
metric $g_{i j}$
      \begin{eqnarray} \label{R}
      {R_{i j}}_{|_{y=0}}&=&{{\widetilde R}_{i j |_{y=0}}}
      -{\frac{{g_{tt}}_{,i j}}{2g_{tt }}}_{|_{y=0}}
      +{\frac{{g_{tt}}_{,i} \, {g_{tt}}_{,j}}{4{g_{tt}}^2}}_{|_{y=0}},
      \
      {R^t_{t}}_{|_{y=0}}=
      {\frac{\delta^{i j}{g_{tt}}_{,i j}}{2 g_{tt}}}_{|_{y=0}}
      -{\frac{\delta^{i j}{g_{tt}}_{,i} \, {g_{tt}}_{,j}}{4 g_{tt}^2}}_{|_{y=0}},
      \nonumber \\
      {R^t_{i}}&=&0, \ {R}_{|_{y=0}}={{\widetilde R}_{|{y=0}}}
      -{\frac{\delta^{i j}{g_{tt}}_{,i j}}{g_{tt}}}_{|_{y=0}}
      +{\frac{\delta^{i j}{g_{tt}}_{,i} \, {g_{tt}}_{,j}}{24 g_{tt}^2}}_{|_{y=0}}.
      \end{eqnarray}
Defining $\overline{G} (y^i)$ by relation
        \begin{equation} \label{Gw}
        \overline G(y^i)=\sqrt{g^{(3)}} G_{\mbox{\tiny E}}(y^i)
        \end{equation}
and retaining in (\ref{beq}) only terms with coefficients involving
two derivatives of the metric or fewer one finds that $\overline G
(y^i)$ satisfies the equation
        \begin{eqnarray}
        &&\delta^{i j}\frac{\partial^2 \overline G}{\partial y^i \partial y^j}
        -m_{\mbox{\tiny ef}}^2 \overline G
        +\delta^{i j} \frac{{g_{tt}}_{,i}}{2 {g_{tt}}}
        \frac{\partial \overline G}{\partial y^j}
        +\delta^{i j}\left(\frac{{g_{tt}}_{,i k}}{2 {g_{tt}}}
        -\frac{{g_{tt}}_{,i} \, {g_{tt}}_{,k}}{2 {g_{tt}}^2}
        \right)y^k \frac{\partial \overline G}{\partial y^j}
        \nonumber \\ &&
         +{\widetilde R}_{\phantom{i} k \phantom{j} l }^{i \phantom{k} j} \ \frac{y^k y^l}{3}
        \frac{\partial^2 \overline G}{\partial y^i \partial y^j}
        +\delta^{i j}\left(\frac{{g_{tt}}_{,i j}}{2 {g_{tt}}}
        -\frac{{g_{tt}}_{,i} \, {g_{tt}}_{,j}}{2 {g_{tt}}^2}
        \right) \overline G
        +\frac{\widetilde R}{3}\, \overline G = -\delta^{(3)}(y),
        \end{eqnarray}
where the coefficients here and below are evaluated at $y^i=0$
(i.e. at the point ${x^i}'$).
Let us represent the Green function $\overline G$ as a series
        \begin{equation} \label{Gy}
        \overline G(y^i)=\overline G_0(y^i) + \overline
        G_1(y^i) + \overline G_2(y^i)+\dots,
        \end{equation}
where $\overline G_a(y^i)$ incorporates a geometrical coefficients involving
$a$ derivatives of the metric at point $y^i=0$. These functions satisfy the set of equations
        \begin{equation}
        \delta^{i j}\frac{\partial^2 \overline G_0}
        {\partial y^i \partial y^j}-m_{\mbox{\tiny ef}}^2 \overline
        G_0=-\delta^{(3)}(y),
        \end{equation}
        \begin{equation}
        \delta^{i j}\frac{\partial^2 \overline G_1}
        {\partial y^i \partial y^j}-m_{\mbox{\tiny ef}}^2 \overline G_1
        +\delta^{i j}\frac{{g_{tt}}_{,i}}{2{g_{tt}}}\frac{\partial \overline G_0}
        {\partial y^j}=0,
        \end{equation}
        \begin{eqnarray} \label{G2}
        &&\delta^{i j}\frac{\partial^2 \overline G_2}
        {\partial y^i \partial y^j}-m_{\mbox{\tiny ef}}^2 \overline G_2
        +\delta^{i j} \frac{{g_{tt}}_{,i}}{2 {g_{tt}}}
        \frac{\partial \overline G_1}{\partial y^j}
        +\delta^{i j}\left(\frac{{g_{tt}}_{,i k}}{2 {g_{tt}}}
        -\frac{{g_{tt}}_{,i} \, {g_{tt}}_{,k}}{2 {g_{tt}}^2}
        \right)y^k \frac{\partial \overline G_0}{\partial y^j}
        \nonumber \\ &&
         +{\widetilde R}_{\phantom{i} k \phantom{j} l}^{i\phantom{k}j} \ \frac{y^k y^l}{3}
        \frac{\partial^2\overline  G_0}{\partial y^i \partial y^j}
        +\delta^{i j}\left(\frac{{g_{tt}}_{,i j}}{2 {g_{tt}}}
        -\frac{{g_{tt}}_{,i} \, {g_{tt}}_{,j}}{2 {g_{tt}}^2}
        \right) \overline G_0
        +\frac{\widetilde R}{3}\, \overline G_0=0.
        \end{eqnarray}
The function $\overline G_0(y^i)$ obeys the condition
        \begin{equation}
        {\widetilde R}_{\phantom{i} k \phantom{j} l}^{i\phantom{k}j}\, y^k y^l  \frac{\partial^2
        \overline G_0}{\partial y^i \partial y^j} -
        {\widetilde R}^i_{j}\, y^j \frac{\partial \overline G_0}
        {\partial y^i}=0,
        \end{equation}
since $\overline G_0(y^i)$ should be the function of $\delta_{i
j} y^i y^j$ only. Therefore Eq. (\ref{G2}) may be rewritten
        \begin{eqnarray} \label{G2n}
        \delta^{i j}\frac{\partial^2 \overline G_2(y^i)}
        {\partial y^i \partial y^j}-m_{\mbox{\tiny ef}}^2 \overline G_2(y^i)
        &+&\delta^{i j} \frac{{g_{tt}}_{,i}}{2 {g_{tt}}}
        \frac{\partial \overline G_1}{\partial y^j}
        +\left[\frac13 {\widetilde R}^i_{k}+
        \delta^{i j}\left(\frac{g_{tt,j k}}{2 {g_{tt}}}
        -\frac{g_{tt,j}{g_{tt,k}}} {2 {g_{tt}}^2}
        \right) \right]  \, y^k
        \frac{\partial \overline G_0}{\partial y^i }
        \nonumber \\
        &+&\delta^{i j}\left(\frac{{g_{tt}}_{,i j}}{2 {g_{tt}}}
        -\frac{{g_{tt}}_{,i} \, {g_{tt}}_{,j}}{2 {g_{tt}}^2}
        \right) \overline G_0
        +\frac{\widetilde R}{3}\, \overline G_0=0.
        \end{eqnarray}
Let us introduce the local momentum space associated with the
point $y^i=0$ by making the 3-dimensional Fourier transformation
        \begin{equation} \label{ms}
        \overline G_a(y^i)= \int \!\!\!\!\int \limits_{-\infty}
        ^{+\infty}\!\!\!\!\int \frac{dk_1 dk_2 dk_3}{(2 \pi)^3}
        e^{i k_i y^i} \overline G_a(k^i).
        \end{equation}
It is easy to check that
        \begin{equation} \label{G0k}
        \overline G_0(k^i)=\frac{1}{k^2+m_{\mbox{\tiny ef}}^2},
        \end{equation}
        \begin{equation} \label{G1k}
        \overline G_1(k^i)=i\frac{\delta^{i j}{g_{tt}}_{,i}k_j}
        {2{g_{tt}}(k^2+m_{\mbox{\tiny ef}}^2)^2},
        \end{equation}
        \begin{equation} \label{G2k}
        \overline G_2(k^i)=
        \frac{\displaystyle k_i k_j \delta^{i k} \delta^{j \, l}
        \left( \frac23 {\widetilde R}_{k \, l}
        + \frac{ {g_{tt}}_{,k \, l}}{{g_{tt}}}
        -\frac{5{g_{tt}}_{,k} \, {g_{tt}}_{,l}}{4 {g_{tt}}^2}\right)}
        {(k^2+m_{\mbox{\tiny ef}}^2)^3},
        \end{equation}
where $k^2=\delta^{i j}k_i k_j$. Taking into account Eqs. (\ref{ms}), (\ref{G0k}), (\ref{G1k}), (\ref{G2k})  in (\ref{Gy}) and integrating over $k_1, k_2, k_3$ we obtain
        \begin{eqnarray}
        \overline G_0(y^i) + \overline G_1(y^i) + \overline G_2(y^i)&=&
         \frac{\exp (-m_{\mbox{\tiny ef}} y)}{8 \pi }
         \left\{\frac2y-\frac{{g_{tt}}_{,i}}{2{g_{tt}}} \frac{y^i}{y}
         +\frac{1}{m_{\mbox{\tiny ef}}} \left[\delta^{i j}
         \left( \frac{{g_{tt}}_{,i j}}{4 {g_{tt}}}
         -\frac{5 \delta^{i j}{g_{tt}}_{,i} \, {g_{tt}}_{,j}}{16 {g_{tt}}^2}\right)
         +\frac{\widetilde R}{6}   \right]
         \right. \nonumber \\ && \left.
        +\left( -\frac{{g_{tt}}_{,i j}}{4 {g_{tt}}}
        +\frac{{g_{tt}}_{,i} \, {g_{tt}}_{,j}}{4 {g_{tt}}^2}
        -\frac{{\widetilde R}_{i j}}{6} \right) \frac{y^i y^j}{ y}\right\},
        \end{eqnarray}
where
        \begin{equation}
        y=\sqrt{\delta_{i j}y^i y^j}.
        \end{equation}
Using the definition of $\overline G(y^i)$ (\ref{Gw}), the expansions
(\ref{exg}), and (\ref{R}) one finds
        \begin{eqnarray} \label{Gwy}
        G_{\mbox{\tiny E}}(y^i)&=& \frac{1}{8 \pi } \left\{\frac2y-\frac{{g_{tt}}_{,i}}{2{g_{tt}}}\frac{y^i}{y}
         -2m_{\mbox{\tiny ef}} +\frac{1}{m_{\mbox{\tiny ef}}}
         \left[\frac{R}{6}+\delta^{i j}\left(\frac{5{g_{tt}}_{,i j}}{12 {g_{tt}}}
         -\frac{19{g_{tt}}_{,i} \, {g_{tt}}_{,j}}{48 {g_{tt}}^2}
         \right) \right]
         \right. \nonumber \\ && \left.
         +m_{\mbox{\tiny ef}}\frac{{g_{tt}}_{,i} \, y^i}{2{g_{tt}} }+m_{\mbox{\tiny ef}}^2 y
         +\left[ \delta^{i j} \left( -\frac{5 {g_{tt}}_{,i j}}{12 {g_{tt}}}
         +\frac{19 {g_{tt}}_{,i} \, {g_{tt}}_{,j}}{48 {g_{tt}}^2} \right)
         -\frac{R}{6}\right] y
         \right. \nonumber \\ && \left.
        +\left( -\frac{{g_{tt}}_{,i j}}{6 {g_{tt}}}
        +\frac{5{g_{tt}}_{,i} \, {g_{tt}}_{,j}}{24 {g_{tt}}^2}
        +\frac{R_{i j}}{6}  \right) \frac{y^i y^j}{ y}
        \right. \nonumber \\ && \left.
        +O\left(\frac{1}{m_{\mbox{\tiny ef}}^2 L^3}\right)+O\left(\frac{y}{m_{\mbox{\tiny ef}} L^3}\right)
        +O\left(\frac{y^2}{L^3}\right) +O\left(\frac{m_{\mbox{\tiny ef}} y^3}{L^3}\right)
        +O\left(m_{\mbox{\tiny ef}}^3 y^2\right)\right\}.
        \end{eqnarray}
In an arbitrary coordinates this expression reads
        \begin{eqnarray} \label{Gwsigma}
        G_{\mbox{\tiny E}}(x^i; {x^i}')
        &=&
        \frac{1}{8 \pi } \left\{\frac{2}{\sqrt{2 \sigma}}
        -2m_{\mbox{\tiny ef}}+\frac{{g_{t't'}}_{,{i}'}\sigma^{{i}'}}{2{g_{t't'}} \sqrt{2 \sigma}}
         +\frac{1}{m_{\mbox{\tiny ef}}}\left(\frac{R}{6}
         +\frac{5{{g_{t't'}}_{,{i}'}}^{;{i}'}}{12 {g_{t't'}}}
         -\frac{19{g_{t't'}}_{,{i}'}{g_{t't'}}^{,{i}'}}{48 {g_{t't'}}^2} \right)
         \right. \nonumber \\ && \left.
         +m_{\mbox{\tiny ef}}^2 \sqrt{2 \sigma}
         -m_{\mbox{\tiny ef}} \frac{{g_{t't'}}_{,{i}'}\sigma^{{i}'}}{2{g_{t't'}} }
         +\left(- \frac{5{{g_{t't'}}_{,{i}'}}^{;{i}'}}{12 {g_{t't'}}}
         +\frac{19{g_{t't'}}_{,{i}'} \, {g_{t't'}}^{,{i}'}}{48 {g_{t't'}}^2}
         -\frac{R}{6}  \right)\sqrt{2 \sigma}
         \right. \nonumber \\ && \left.
        +\left( -\frac{{g_{t't'}}_{,{i}' ; {j}'}}{6 {g_{t't'}}}
        +\frac{5{g_{t't'}}_{,{i}'} \, {g_{t't'}}_{,{j}'}}{24 {g_{t't'}}^2}
        +\frac{R_{{i}' {j}'}}{6}  \right) \frac{\sigma^{{i}'} \sigma^{{j}'}}{ \sqrt{2 \sigma}}
        \right. \nonumber \\ && \left.
        +O\left(\frac{1}{m_{\mbox{\tiny ef}}^2 L^3}\right)
        +O\left(\frac{\sqrt{\sigma}}{m_{\mbox{\tiny ef}} L^3}\right)
        +O\left(\frac{\sigma}{L^3}\right)
        +O\left(\frac{m_{\mbox{\tiny ef}} \sigma^{3/2}}{L^3}\right)
        +O\left(m_{\mbox{\tiny ef}}^2 \sigma \right) \right\},
        \end{eqnarray}
where $\sigma$ is one half the square of the distance between the points $x^i$
and ${x^i}'$ along the shortest geodesic connecting them, $\sigma^{i'}=g^{i'j'}
\partial\sigma/\partial x^{j'}$ and we have taken into account that $g_{tt}$ is the scalar field in $3D$ space
with metric $g_{ij}$. The renormalization  procedure lies in the fact that we
subtract the terms of this expansion which are survived in the limits $m_{\mbox{\tiny ef}} \rightarrow \infty$ and $\sigma \rightarrow 0$
        \begin{equation} \label{GDS}
        A_{(t)}^{sing}(x^i; {x^i}')=-4 \pi e \, G_{\mbox{\tiny E}}^{sing}(x^i; {x^i}')=
        -e \left(\frac{1}{\sqrt{2 \sigma}}
        -m_{\mbox{\tiny ef}}+\frac{g_{t't',i'}\sigma^{i'}}{4{g_{t't'}} \sqrt{2 \sigma}}
        \right).
        \end{equation}
The remaining terms in (\ref{Gwsigma}) give us the possibility
to compute the self-potential and self-force for particle at rest
which is the source of massive field.

\subsection{The Schwarzschild space-time}\label{Sec:BlackHole}

Let us verify above scheme for well-known case of black hole \cite{Vil79,SmiWil80,PfePoi02}.
The electromagnetic potential of a charged particle at rest in the Schwarzschild space-time with metric
\begin{equation*}
ds^2 = - \left(1 - \frac{2M}r\right) dt^2 + \left(1 -
\frac{2M}r\right)^{-1} dr^2 + r^2 (d\theta^2 + \sin^2\theta
d\varphi^2),
\end{equation*}
was obtained by Copson \cite{Cop28} and corrected by Linet in Ref. \cite{Lin76}
\begin{equation}
A_t = -\frac{e}{rr'} \frac{(r - M)(r' -
M) - M^2 \cos \chi}{R} - \frac{e M}{r r'}, \label{CopsonLinet}
\end{equation}
where $R^2 = (r - M)^2 + (r' - M)^2 - 2 (r - M)(r' - M)\cos\chi - M^2
\sin^2\chi$ and $\cos\chi = \cos\theta\cos\theta' + \sin\theta\sin\theta'
\cos (\varphi - \varphi')$.
In the limit of coincided angle variables we obtain
\begin{equation}
A_t = -\frac{e}{|r-r'|}\left(1 - \frac{2M}{r}\right),
\end{equation}
and the tetrad component has the following form
\begin{equation}
A_{(t)} = -\frac{e}{|r-r'|}\sqrt{1 - \frac{2M}{r}}.
\end{equation}
The singular part of this quantity as was shown above reads
\begin{equation}
A_{(t)}^{sing} = -e\left(\frac{1}{\sqrt{2\sigma}} +
\frac{g_{t't',k'} \sigma^{k'}}{4 g_{t't'}\sqrt{2\sigma}}\right).
\end{equation}
where the half-square of the radial geodesic distance in this case is
\begin{equation}
\sigma(r,r') = \frac 12 s(r,r') = \frac 12 \left(\int_{r'}^r \frac{dr}{\sqrt{1-2M/r}}\right)^2.
\end{equation}
Straightforward calculations gives the following expression for divergent part
\begin{equation}
A_{(t)}^{sing} = -\frac{e}{|r-r'|}\sqrt{1 - \frac{2M}{r'}}.
\end{equation}
To renormalize the potential we subtract divergent part and make coincidence limit
\begin{equation}
A_{(t)}^{ren} = \lim_{r'\to r}(A_{(t)} - A_{(t)}^{sing}) = -\frac{eM}{r^2\sqrt{1 - \frac{2M}{r}}}.
\end{equation}
Then we define the self-potential in usual manner (the sign minus is due to covariant index of the $4$-potential)
\begin{equation}\label{Uself}
U^{self} =  -\frac e2 A_{t}^{ren} = \frac{e^2M}{2r^2}.
\end{equation}
With this definition of the self-energy we have classical definition of force (tetrad component) as a minus gradient of the potential,
\begin{equation}\label{Fself}
{\cal F}^{(r)} =  e{F^{r}_\cdot}_{\! t}{}^{self} u^t/\sqrt{-g_{rr}} = -
\partial_rU^{self},
\end{equation}
and the sign of the potential gives the information about attractive or repelling character of the force. By using this expression we recover well-known expression for self-force (tetrad component) in Schwarzschild space-time \cite{Vil79,SmiWil80,PfePoi02}
\begin{equation}\label{sf_blach_hole}
{\cal F}^{(r)} =  \frac{Me^2}{r^3}.
\end{equation}

\section{The self-force}\label{Sec:Self}

Let us proceed here to calculation the self-force in the metric (\ref{metric}).  According with the scheme developed above we have to subtract from $A_{(t)}$ the following expression:
\begin{equation}\label{Asing}
A_{(t)}^{sing} = -e\left(\frac{1}{\sqrt{2\sigma}} +
\frac{g_{t't',k'} \sigma^{k'}}{4 g_{t't'}\sqrt{2\sigma}}\right).
\end{equation}
The curve with constant $t, \theta, \phi$ is the geodesic
line and we have
\begin{equation}
\sigma (\rho,\rho') = \frac 12 s^2 (\rho,\rho') = \frac 12
\left(\int_{\rho'}^{\rho} e^{\alpha(x)/2} dx \right)^2.
\end{equation}
By using this expression we obtain the term to be subtracted ($\rho > \rho' >0$)
\begin{equation}
A_{(t)}^{sing} = -\frac{e}{\rho -\rho'}e^{-\alpha(\rho')/2}.
\end{equation}
Therefore we obtain the expression for renormalized potential
\begin{equation}
A_{(t)}^{ren} = \lim_{\rho'\to\rho}(A_{(t)}^+ - A_{(t)}^{sing}) = \frac{e}{\rho^2 + n^2 - m^2} \frac{m e^{-\alpha/2}}{\tanh\pi b},
\end{equation}
the self-potential
\begin{equation}\label{UselfWH}
U^{self} = - \frac e2 A_{t}^{ren} = -\frac{e^2}{\rho^2 + n^2 - m^2} \frac{m e^{-\alpha}}{2\tanh\pi b}
\end{equation}
and tetrad component of the self-force
\begin{equation}\label{FselfWH}
{\cal F}^{(\rho)} = -\partial_\rho U^{self} = \frac{e^2}{(\rho^2 + n^2 - m^2)^2}   \frac{m (m-\rho)e^{-\alpha}}{\tanh\pi b}.
\end{equation}

For massless wormhole, $m\to 0$, we recover results obtained in Ref.
\cite{KhuBak07}. Far from the wormhole's throat we obtain
\begin{equation}\label{ratios}
\frac{U_{\rho\to +\infty}^{self}}{U_{\rho\to -\infty}^{self}} = \left| \frac{{\cal F}^{(\rho)}_{\rho\to +\infty}}{{\cal F}^{(\rho)}_{\rho\to -\infty}}\right| = e^{2\pi b}.
\end{equation}
The self-force equals to zero at sphere of minimal square and the extreme
of the force is at the points $\rho = m \pm n/\sqrt{3}$.

\begin{figure}[ht]
\begin{center}
\epsfxsize=8truecm\epsfbox{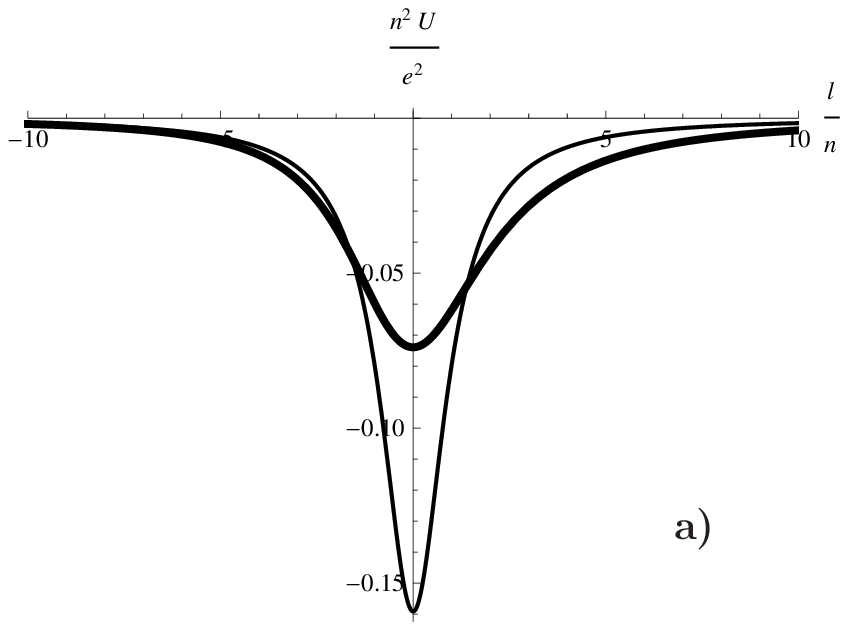}\epsfbox{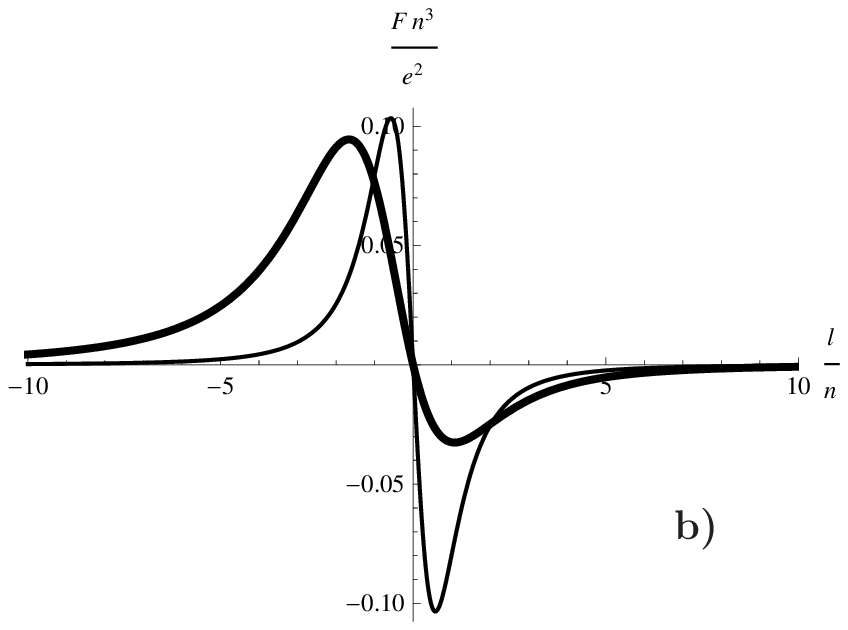}
\end{center}
\caption{The self-energy (a) and the self-force (b) for massless case (thin
curves)  and for massive wormhole case (thick curves) for $m/n=0.7$. The zero
value of the $l$ corresponds to the sphere of the minimal
square in both cases.}\label{fig:self}
\end{figure}

To compare our results with massless case we should plot pictures in the same coordinates. We plot the self-energy and the self-force as a function of the  proper radial coordinate $l$, which is
\begin{equation}
l = \int_m^\rho e^{\alpha (x)/2}dx.
\end{equation}
It is merely $\rho$ for massless case. The modulo of this coordinate is the
proper distance from the sphere of the minimal square. The self-energy and the tetrad component of the self-force are shown in the Fig. \ref{fig:self}. We
observe that the parameter of the mass, $m$, brakes  the symmetry of the figure.
The extrema of the self-force are at the points $\rho = m \pm n/\sqrt{3}$.

Let us analyze our formulas for $n \approx m$. For $m\to n$ we have
\begin{equation}
\alpha = \left\{ \begin{array}{ll}
\frac{2n}{\rho}+O\left( \frac{n-m}{\rho} \right)
+O\left( \frac{n^2(n-m)}{\rho^3} \right), & \rho>0,
\\
\pi \sqrt{\frac{2n}{n-m}}+O\left( \frac{n}{\rho} \right), & \rho < 0 ,
\end{array}\right.
\end{equation}
and therefore the self-energy and self-force tend to zero for $\rho < 0$.
But they have the form showed in Fig. \ref{fig:self} as the functions of coordinate $l$
because the coordinate $l$ tends to minus infinity for $\rho = 0$.
The self-energy at its extreme in this case tends to
zero as
\begin{equation}
U(\rho = 0) \approx -\frac{e^2}{4(n-m)}e^{-\pi\sqrt{\frac{n}{2(n-m)}}}.
\end{equation}
But the self-energy at extremes tends to finite values
\begin{eqnarray}
\lim_{m\rightarrow n}{\cal
F}^{(\rho)}\left(\rho=m+\frac{n}{\sqrt{3}}\right) &=&
\frac{e^2}{n^2}\left(9 - 21 \frac{\sqrt{3}}{4}\right)
e^{-3+\sqrt{3}} \approx - \frac{e^2}{n^2} 0.026\ldots\allowdisplaybreaks\\
\lim_{m\rightarrow n}{\cal F}^{(\rho)}\left(\rho=m-\frac{n}{\sqrt{3}}\right) &=&
\frac{e^2}{n^2}\left(9 + 21 \frac{\sqrt{3}}{4}\right)
e^{-3-\sqrt{3}} \approx + \frac{e^2}{n^2} 0.159\ldots ,
\end{eqnarray}
and the positions of the extremes are $l/n = - 2.76\ldots$ and $l/n = 1.27\ldots$.

\section{Conclusion}\label{Sec:Conclusion}

In the paper we considered in detail the self-interaction on an electrically
charged  particle at rest in a space-time of wormhole the metric of which is a solution to the Einstein-scalar field equations in the case of a phantom massless scalar field.
The main aim of the
calculations is to reveal the role of the massive parameter, $m$, of wormhole in
the self-force. The space-time of massive wormhole is taken in the form
(\ref{metric}) which was obtained in Refs. \cite{Bron73,Ell73}. The parameter
of mass reveals itself in gravitational force acting on the test particle and it
is positive in the one part ($\rho>0$) of the wormhole space-time and negative
in the part with $\rho<0$.

The space-time under consideration has no ultrastatic form, the component $g_{tt} \not = -1$.
In this case we developed the renormalization procedure for tetrad component
of the $4$-potential and found the terms which have to be subtracted  (\ref{GDS})
from the potential for renormalization. The application of this scheme gives
well-known result (\ref{sf_blach_hole}) for self-force of particle at rest
in the Schwarzschild space-time.

The definition the self-potential in the form (\ref{Uself}) gives us the standard expression
for tetrad component of self-force (\ref{Fself}) as minus gradient the self-potential
\begin{equation}
{\cal \mathbf{F}}^{self} = - \nabla U^{self},
\end{equation}
and the sign of the potential marks the repelling or attractive
character of the self-force as in classical mechanics.

For the space-time under consideration we found simple and close
expressions  for self-potential (\ref{UselfWH}) and self-force
(\ref{FselfWH}). The space-time has no symmetry for changing $\rho
\to -\rho$. The self-force and the self-energy in its turn have no
the symmetry, too. More precisely, the ratio the self-forces or
self-potentials at infinity and minus infinity equals to $\exp
(2\pi m/\sqrt{n^2-m^2})$ (\ref{ratios}). The self-force attracts
particles to the throat and it falls down as $\rho^{-3}$ far from
the throat. The extrema of the self-force are at the points $\rho=
m\pm n/\sqrt{3}$. The self-force and the self-energy have the form
plotted in Fig. \ref{fig:self}. We observe that the self-force as a
function of invariant proper radial coordinate $l$ has the form similar to that in massless wormhole case. The parameter of mass changes 
asymmetrically the form of the curve and it saves the attractive
character of the force.

\acknowledgements

This work was supported in part by the Russian Foundation for
Basic Research grants No. 08-02-00325.

\end{document}